\documentclass[epj,twocolumn]{webofc}
\usepackage[varg]{txfonts}   
\usepackage{multicol}
\usepackage{color}
\usepackage[normalem]{ulem}

\woctitle{Powders \& Grains 2017}
\begin{document}
\title{An experimental investigation of the force network ensemble}
%
%

\author{\firstname{Jonathan E.} \lastname{Kollmer}\inst{1}\fnsep\thanks{\email{jekollme@ncsu.edu}} \and
        \firstname{Karen E.} \lastname{Daniels}\inst{1}
}

\institute{Department of Physics, North Carolina State University, Raleigh, North Carolina 27695, USA
          }

\abstract{%
We present an experiment in which a horizontal quasi-2D granular system with a fixed neighbor network is cyclically compressed and decompressed over 1000 cycles.  We remove basal friction by floating the particles on a thin air cushion, so that particles only interact in-plane.  As expected for a granular system, the applied load is not distributed uniformly, but is instead concentrated in force chains which form a network throughout the system. To visualize the structure of these networks, we use particles made from photoelastic material. The experimental setup and a new data-processing pipeline allow us to map out the evolution subject to the cyclic compressions. We characterize several statistical properties of the packing, including the  probability density function of the contact force, and compare them with theoretical and numerical predictions from the force network ensemble theory. }
\maketitle
\section{Introduction \label{introduction}}

Although the positions of particles in a jammed granular system are fixed to a specific geometrical configuration, the particle positions alone are not sufficient to determine the force network that carries the load on that packing. As such, an underdetermined mechanical system, there are many ways in wich force and torque balance on each particle can be statistfied for any given packing geometry and boundary conditions \cite{Snoeijer2004,Tighe2010}. 
There are both stable configurations, and unstable configurations and initially stable granular systems can evolve into catastrophic failure.
While two packings might have the same occupied volume or internal pressure they might have vastly different bulk material properties \cite{Dagois-Bohy2012}.
For frictional granular system it remains an open question to determine whether the statistics due to the stress state and the volume of the system can be decoupled from each other.

\begin{figure}[h]
\centering
\includegraphics[width=\linewidth]{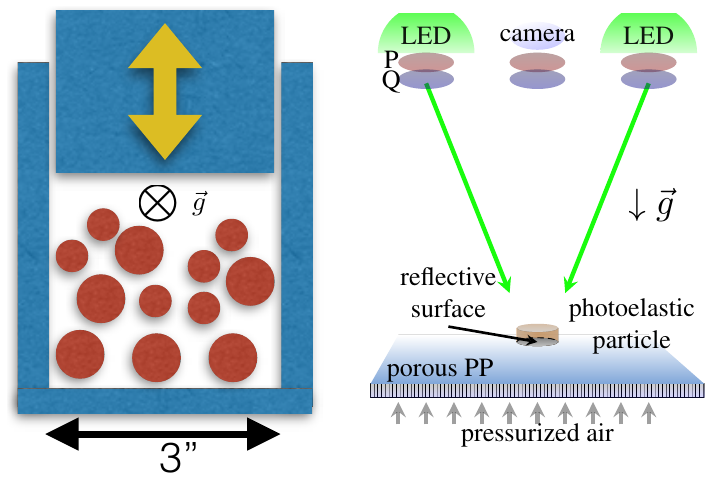}
\caption{Experimental setup. Left: Particles are placed into a piston that cyclically compresses them.  Right: The whole setup is horizontal and floated on an air cushion to eliminate the influence of gravity and basal friction. A camera and an unpolarized red light source are mounted overhead to image and track the particles. The particles are made from photoelastic material, providing visual access to their internal stress field when illuminated with polarized green light. The red and green color channels of the camera can later be seperated to process particle positions and force information from a single color image. Right subfigure adapted from \cite{Puckett-thesis}. 
}
\label{fig-setup}       
\end{figure}

To make predictions for the physical behavior of granular systems, tools and concepts from statistical physics have become widely used \cite{Bi2015}, but  what is the correct ensemble to describe jammed granular packings? 
To get more insight into these jammed packings one needs to look at the structure of force networks that form in loaded granular packings.
When the packing is subjected to external load, not all particles share the load equally but the forces are highly localized into force chains.
In this work, we present an experiment to look at the distribution of forces in a loaded granular system, 
while disentangling the effects of configuration from other influences. This is of interest to compare these distributions to predictions from the Force Network Ensemble theory \cite{Snoeijer2004}.

\subsection*{The Force Network Ensemble}

The Force Network Ensemble (FNE) is a concept introduced by Snoeijer et al. \cite{Snoeijer2004}  
in which they use an ensemble approach for examining the force distribution in static granular packings. Since forces in
 fixed granular packings are typically underdetermined, the ensemble averages over 
 all microscopic variations of a packing, an approach that goes back to Edwards  \cite{Edwards1989}.  The FNE predicts, among other things \cite{Tighe2010},
 a fitnite value for $P(F)$  as $F \rightarrow 0$ and a faster than exponential decay of the probability density function (PDF) of the contact forces at large forces, depending
 on the dimensionality of the packing \cite{Snoeijer2004}. For a two dimensional system, $P(F)$ is predicted to have a gaussian tail. For a review see \cite{Tighe2010} (and references therein)
 where it is also discussed that the peak in the PDF should vanish for anisotropic stresses.
Saithoh et al. \cite{Saitoh2015}, using molecular dynamics simulations to determine transition rates for contact
changes, recently also found a master equation that describes the PDF of forces in soft particle packings.


\section{An Experimental Approach}

While a number of experimental works exist, e.g. \cite{Corwin2005, Howell1999},  PDFs of contact forces are most often 
produced from numerical simulations, including a recent pair of papers by Pugnaloni \cite{Pugnaloni2016} and Kondic \cite{Kondic2016} 
where they study the structure of force networks in tapped particulate systems of disks superimposed by a gravity force.
However, in most experimental studies the external load that probes the force network is not the only force applied
 to the system, there is additionally a load superimposed by gravity, or basal friction \cite{Kovalcinova2016}. Further, there is only few experiments \cite{Puckett2013} probing granular ensembles. 
 
In this manuscript, we present an experiment that is designed to enumerate how many force configurations of 
a single hyperstatic granular arrangement are practically accessible, while at the same time keeping
the external load the only force that is beeing applied to the system. 
To achieve this we prepare a horizontal quasi 2D granular system that is floated on a gentle air cushion,
thereby generating an effectively gravity free system without basal friction \cite{Puckett-thesis}. 
The particles are confined by a piston that can apply an uniaxial load to the packing.
A schematic drawing of the experimental setup is detailed in Fig.~\ref{fig-setup}. 
By cyclically loading and unloading the packing in a way that 
will not change the particle configuration (no neighbors changes), the system 
cycles trough many contact force configurations due to microscopic changes of the exact contact point. 
For the experiments done here, we compress the packing in steps of constant volume ($\Delta V = 0.002869 V_{initial}$)
and the compression steps are applied quasistatically over 20 substeps of $\Delta x = 0.01 ~\text{mm}$. 
The initial volume $V_{initial}$ was chosen to be close to the onset of jamming, and the final volume so that the mean contact force rises by more than
a factor of 3. We performe the experiment with several random configuration of 29 particles
of two different radii ($r_1 = 5.5 \text{mm}$ and $r_2 = 7.6 \text{mm}$) to prevent crystallization.


In order to extract force information from the experiment, the particles are made of photoelastic material (Vishay PSM-4),
which will shift the polarization of light that is shined through it as a function of the applied load. 
A model of the force modulated light intensity can then be fitted to camera images of the particles
\cite{Daniels2016, Puckett2013}.

\section{Results \label{results}}

\begin{figure*}
\centering
\includegraphics[width=0.9\linewidth]{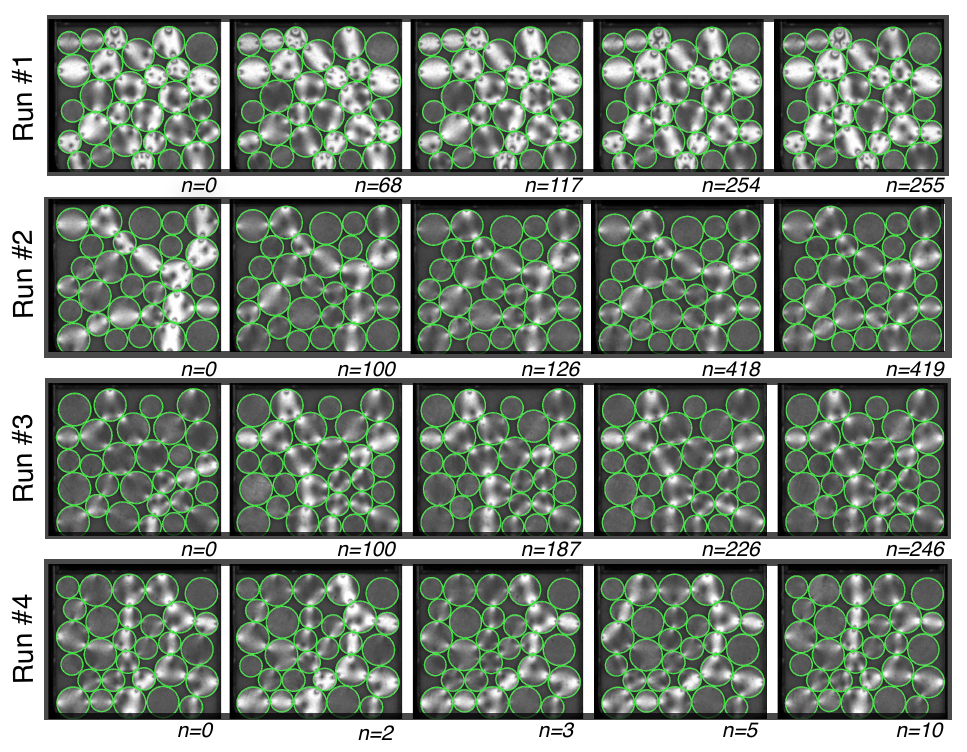}
\caption{Snapshots after $n$ loading cycles for four different initial configurations (Runs \#1- 4). The force network is different for each configuration and for the same configuration fluctuates around several preferred sates.}
\label{fig-loadcycling}       
\end{figure*}

When we start the load cycling we observe that even after an initial annealing period
there are variations in the force network, (see Fig.~\ref{fig-loadcycling}), while the particle configuration 
is unchanged. This validates our experimental approach and allows us to probe the nature of the FNE.  

\begin{figure}[]
\centering
\includegraphics[width=0.9\linewidth]{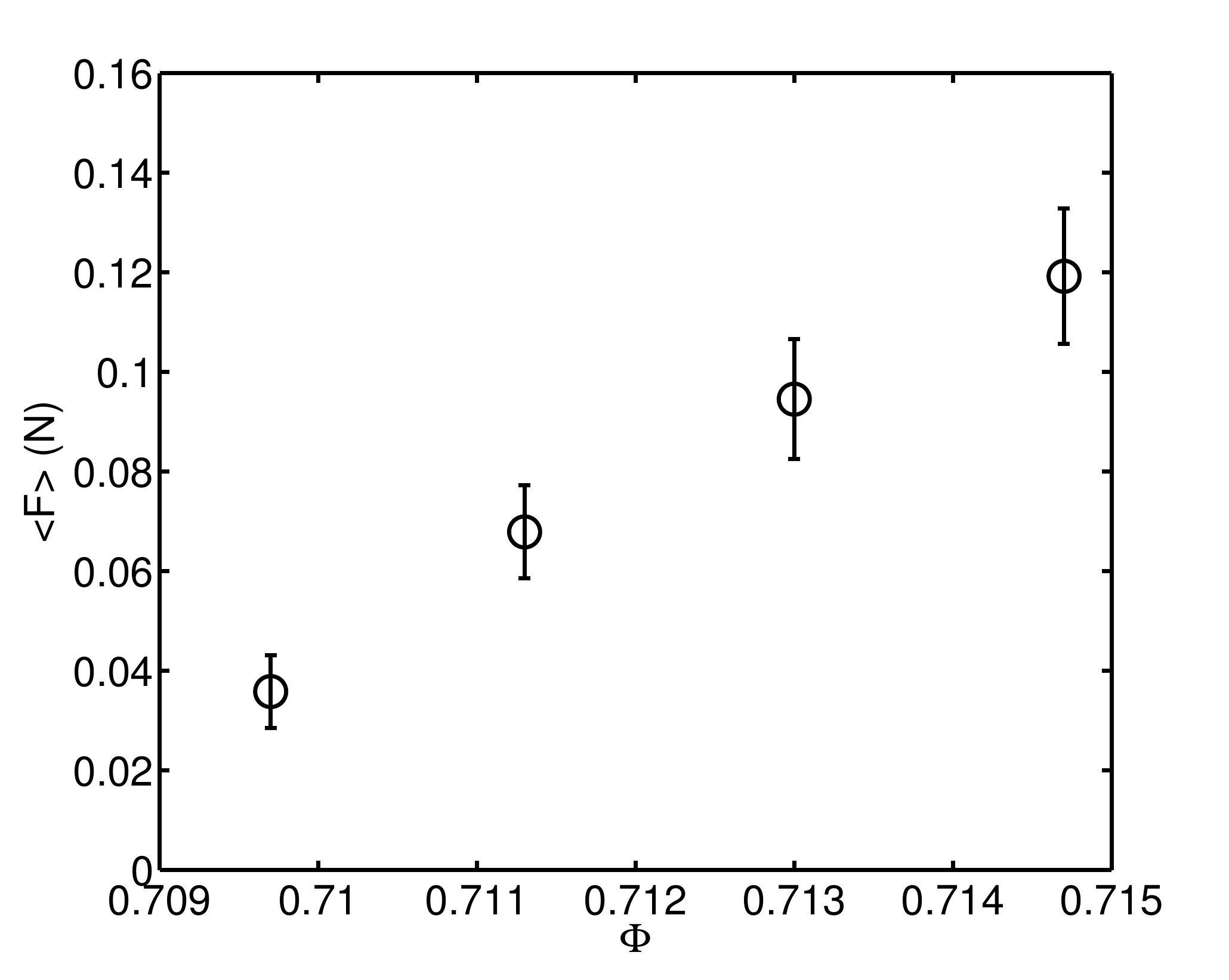}
\caption{Average contact force $F$, averaged over all repetitions of a single experiment  for each loading step (packing fraction). As can be expected $F$ increases as compression increases, and the standard deviation (errorbars) increases with increasing load. }
\label{fig-compression}       
\end{figure}

We run the experiment for $\approx 1000$ cycles and observe strong fluctuations in the contact number, determined by a minimum threshold force ($F_{th} > 0.01$~N)
and in the number of load bearing particles (the number of particles with one ore more contacts above the threshold force).
Figure~\ref{fig-compression} shows the average contact force
 $$F = \Big \langle \sqrt{F_N^2+F_T^2} ~~ \Big \rangle$$
and its standard deviation over all cycles, as a function of the applied compression.
We see that the standard deviations in $F$ grow with the applied compression. For higher packing fractions, the intervals given by the STD($F$) around $F$ begin to overlap for consecutive compression steps. 

\begin{figure}

\centering
\includegraphics[width=0.9\linewidth]{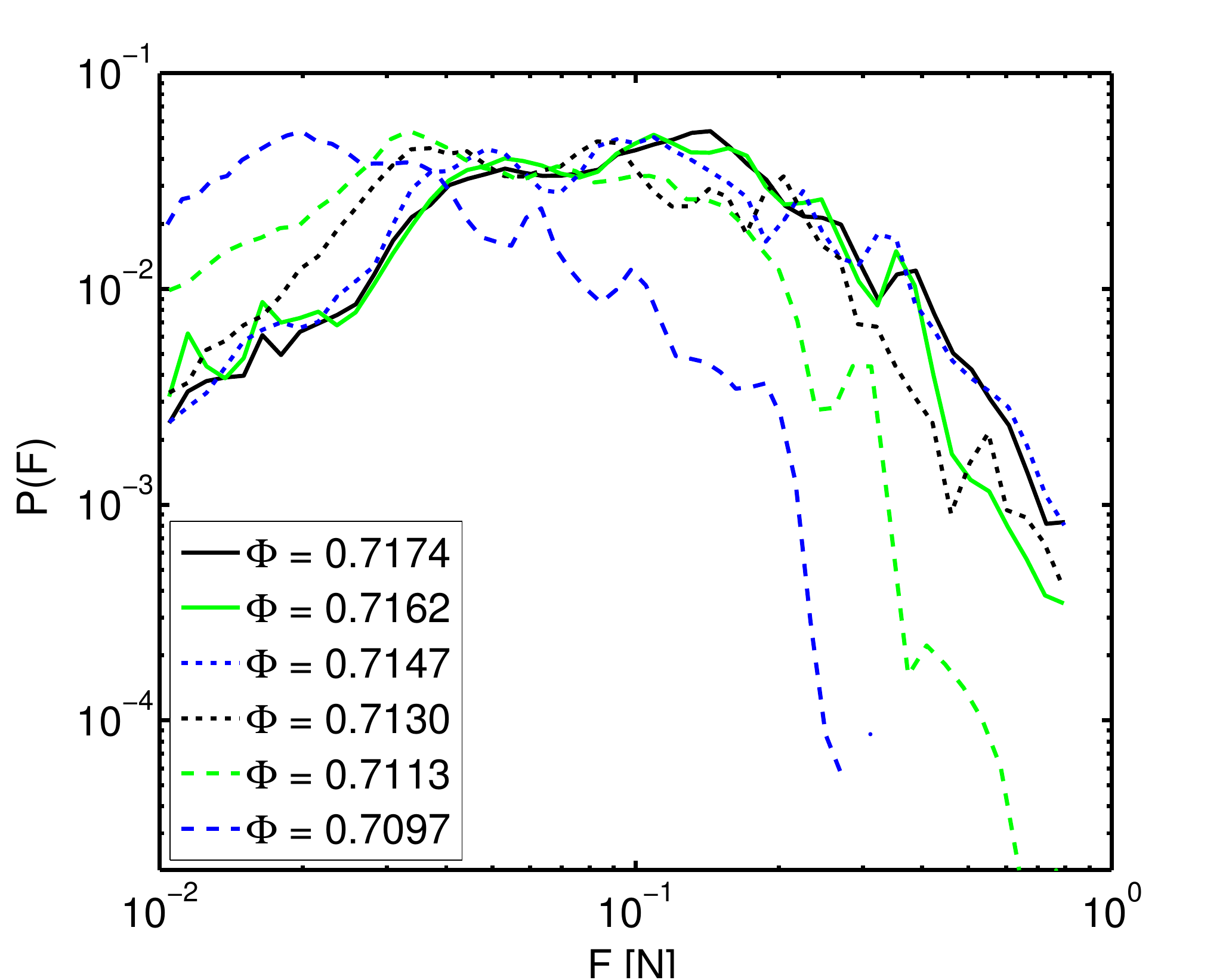}
\includegraphics[width=0.9\linewidth]{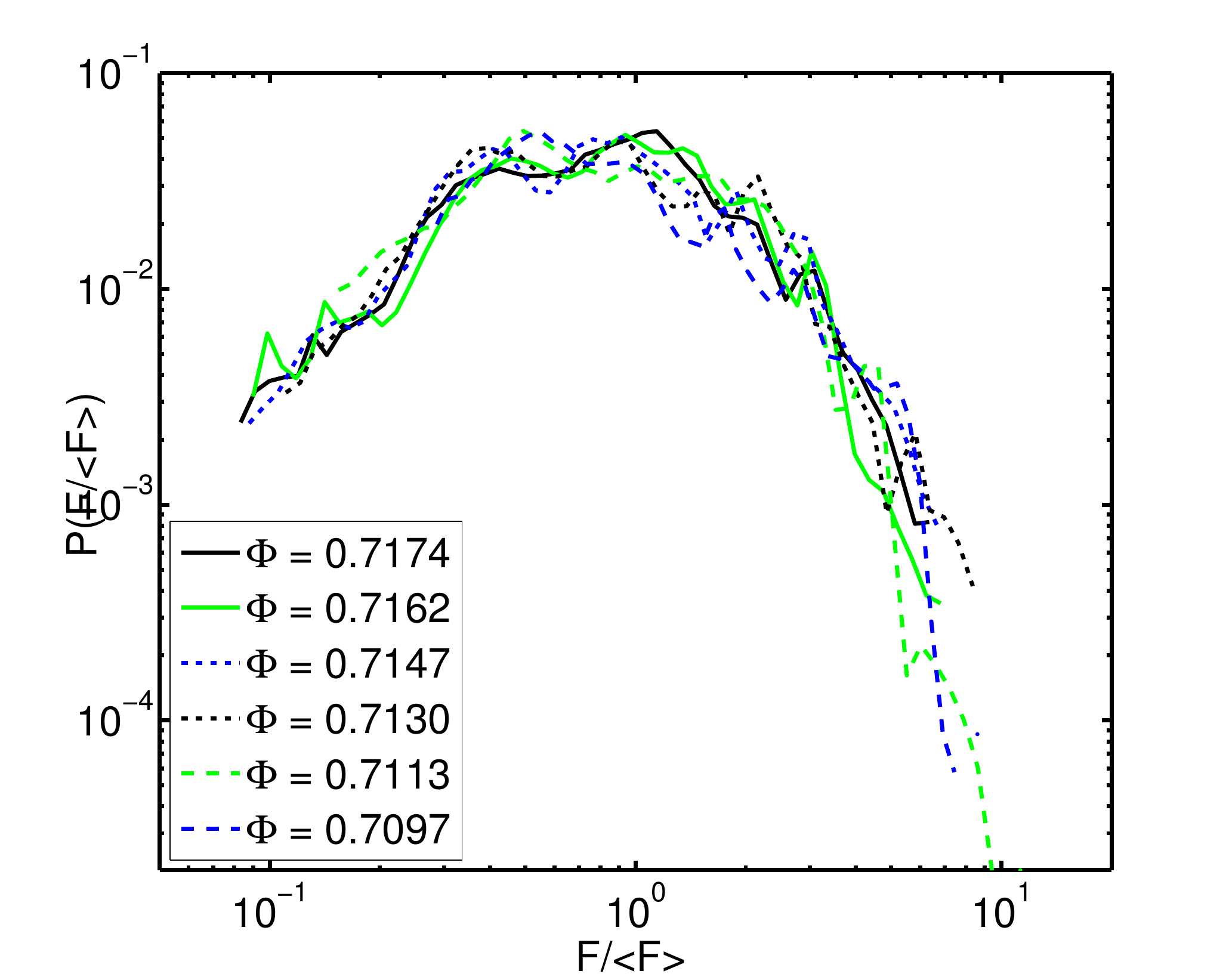}
\caption{Left: Contact force probability distribution function for each loading step. The peak shifts to higher forces with increased load. for smaller loads the PDF seems to have a shoulder, this could be from a strong and weak force network or from two preferred configurations that the systems osciallated inbetween. Right: Contact force PDF for each loading step collapse when normalized by the average contact force of the corresponding loading step. In the limit of large forces the PDFs decay faster than exponential. }
\label{fig-fpdf}       
\end{figure}

Figure~\ref{fig-fpdf} shows the PDFs of the contact forces for 5 compression steps over all cycles.  The PDF exhibits a strong peak, decaying in the limit of  both low and high forces.   These results are qualitatively similar to the predictions by FNE theory: most notably  a stronger than exponential decay in $P(F)$.
We also see the peak in $P(F)$, along with the average contact force, move towards higher values as the system gets compressed stronger. In fact, the PDFs collapse onto a single curve when normalized by the average contact force at the corresponding compression step.  Although we perform an experiment with anisotropic loading, we nonetheless identify a peak in the distribution, a feature that is suggested to vanish for anisotropic loads \cite{Tighe2010}. 
For small forces we find, $P(F)$ to rise exponentially, as approximately $F^{3/2}$ . Wyart  \cite{Wyart2012}, showed that the exponent is determined by the pair distribution function g(r). However, in our small system with a fixed particle configuration, g(r) is undersampled, so the question arises what the exponent is set by here. 

\section{Conculsions and Outlook}
We have designed an experiment that can explore the Force Network Ensemble of a two dimensional granular packing, while excluding forces other than the applied load. We find that we can qualitatively reproduce some of the features in contact force distribution predicted by the FNE theory. 

Due to the limited number of cycles and initial packing configurations explored here, it is not immediately clear whether some features we observed in the contact force PDFs are due to sample size or the specific initial configuration. Repeating the experiment not only for more compression cycles but also many different packing configurations, would allow us to probe ergodic effects by contrasting the time average statistics to the ensemble average statistics.

Furthermore, future work should try to identify whether the are several subpopulations of networks which differ in their contact force PDFs. These subpopulations can be found by using tools from network theory such as community detection \cite{Bassett2015}. More generally, the data generated by this experiment will be useful in relating force network features to macroscopic packing properties \cite{Giusti2016,Papadopoulos2016}.

\section*{Acknowledgements and References}
We gratefully acknowledge James Puckett for the design and construction of the air table on which the apparatus is based, and for the inspiration for the new parallelized version of the contact-force code. This research was suppored by the James S. McDonnell Foundation and the NSF through grants DMR-0644743 and DMR-1206808.

\bibliography{ensembles}

\begin{thebibliography}{18}

\bibitem{Snoeijer2004}
J.H. Snoeijer, T.J.H. Vlugt, M.~van Hecke, W.~van Saarloos, Physical Review
  Letters \textbf{92}, 54302 (2004)

\bibitem{Tighe2010}
B.P. Tighe, J.H. Snoeijer, T.J.H. Vlugt, M.~van Hecke, Soft Matter \textbf{6},
  2908 (2010)

\bibitem{Dagois-Bohy2012}
S.~Dagois-Bohy, B.P. Tighe, J.~Simon, S.~Henkes, M.~van Hecke, Phys. Rev. Lett.
  \textbf{109}, 095703 (2012)

\bibitem{Puckett-thesis}
J.G. Puckett, Phd thesis, North Carolina State University (2012),
  \texttt{http://www.lib.ncsu.edu/resolver/1840.16/7998}

\bibitem{Bi2015}
D.~Bi, S.~Henkes, K.~Daniels, B.~Chakraborty, Annual Review of Condensed Matter
  Physics \textbf{6}, 63 (2015)

\bibitem{Edwards1989}
S.F. Edwards, R.B.S. Oakeshott, Physica A \textbf{157}, 1080 (1989)

\bibitem{Saitoh2015}
K.~Saitoh, V.~Magnanimo, S.~Luding, Soft Matter \textbf{11}, 1253 (2015)

\bibitem{Corwin2005}
E.I. Corwin, H.M. Jaeger, S.R. Nagel, Nature \textbf{435}, 1075 (2005)

\bibitem{Howell1999}
D.~Howell, R.P. Behringer, C.~Veje, Physical Review Letters \textbf{82}, 5241
  (1999)

\bibitem{Pugnaloni2016}
L.A. Pugnaloni, C.M. Carlevaro, M.~Kram\'ar, K.~Mischaikow, L.~Kondic, Phys.
  Rev. E \textbf{93}, 062902 (2016)

\bibitem{Kondic2016}
L.~Kondic, M.~Kram\'ar, L.A. Pugnaloni, C.M. Carlevaro, K.~Mischaikow, Phys.
  Rev. E \textbf{93}, 062903 (2016)

\bibitem{Kovalcinova2016}
L.~Kovalcinova, A.~Goullet, L.~Kondic, Physical Review E \textbf{93}, 042903
  (2016)

\bibitem{Puckett2013}
J.G. Puckett, K.E. Daniels, Physical Review Letters \textbf{110}, 058001 (2013)

\bibitem{Daniels2016}
K.E. Daniels, J.E. Kollmer, J.G. Puckett, arXiv:1612.03525  (2016)

\bibitem{Wyart2012}
M.~Wyart, Phys. Rev. Lett. \textbf{109}, 125502 (2012)

\bibitem{Bassett2015}
D.S. Bassett, E.T. Owens, M.A. Porter, M.L. Manning, K.E. Daniels, Soft Matter
  \textbf{11}, 2731 (2015)

\bibitem{Giusti2016}
C.~Giusti, L.~Papadopoulos, E.T. Owens, K.E. Daniels, D.S. Bassett, Physical
  Review E \textbf{94}, 032909 (2016)

\bibitem{Papadopoulos2016}
L.~Papadopoulos, J.G. Puckett, K.E. Daniels, D.S. Bassett, Physical Review E
  \textbf{94}, 032908 (2016)

\end{thebibliography}


\end{document}